\documentclass[prl,aps,showpacs,twocolumn]{revtex4}
\usepackage{graphics,bm}
\usepackage{amssymb}
\usepackage{epsfig}
\usepackage{epsf}
\usepackage[usenames]{color}

\begin{document}

\title{Spin drag in noncondensed Bose gases}

\author{R.A. Duine}
%\email{R.A.Duine@uu.nl} \homepage{http://www.phys.uu.nl/~duine}

\author{H.T.C. Stoof}
%\email{H.T.C.Stoof@uu.nl} \homepage{http://www.phys.uu.nl/~stoof}

\affiliation{Institute for Theoretical Physics, Utrecht
University, Leuvenlaan 4, 3584 CE Utrecht, The Netherlands}
\date{\today}

\begin{abstract}
We show how time-dependent magnetic fields lead to spin motive
forces and spin drag in a spinor Bose gas. We propose to observe
these effects in a toroidal trap and analyze this particular
proposal in some detail. In the linear-response regime we define a
transport coefficient that is analogous to the usual drag
resistivity in electron bilayer systems. Due to Bose enhancement
of atom-atom scattering, this coefficient strongly increases as
temperature is lowered. We also investigate the effects of
heating.
\end{abstract}


\maketitle

% definitions
\def\bx{{\bf x}}
\def\bk{{\bf k}}
\def\half{\frac{1}{2}}
\def\args{(\bx,t)}

 {\it Introduction: Coulomb drag and spin Coulomb drag}
--- Understanding electronic transport \cite{rammerbook} is one of
the most important goals of condensed-matter physics. Indeed,
materials are often characterized according to their transport
properties. Furthermore, transport measurements provide important
physical information. For example, the temperature dependence of
transport coefficients, like resistivity and conductivity, contain
information on the elementary excitations and their scattering
mechanisms. Moreover, the magnetic-field dependence allows for
extracting the electronic phase-coherence length.

\begin{figure}
\vspace{-0.5cm}
\centerline{\epsfig{figure=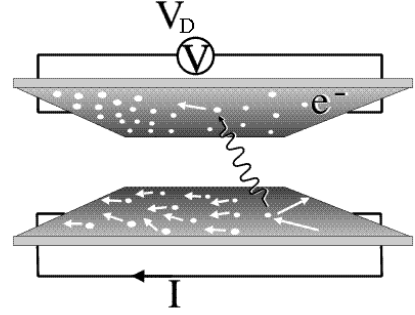,width=7cm}}
 \caption{Illustration of a Coulomb drag measurement:
 a pair of two-dimensional electron gases is separated by a tunnel barrier.
 A current is applied in one of the layers leading to a voltage drop in the other. (Adapted with permission from
 http://www.sp.phy.cam.ac.uk/SPWeb/.)}
 \label{fig:dragexpt}
\end{figure}

Analyzing results of transport measurements is complicated by the
multitude of effects, like electron-electron and electron-phonon
interactions, that contribute. This problem is to a large extent
circumvented in the Coulomb drag measurement of Gramila {\it et
al.} \cite{gramila1991}, illustrated in Fig.~\ref{fig:dragexpt},
that aims at singling out the electron-electron interactions from
the start. In this setup a bilayer of two-dimensional electron
gases is separated by a tunnel barrier. A current $I$ is driven
through the bottom layer that drags along the electrons in the
other layer. In the top layer, an electrochemical potential is
built up that cancels the drag and induces a voltage drop $V_D$,
which results in a drag resistivity $\rho_D = V_D/I$. Originating
from electron-electron interactions, this resistivity usually has
the typical $\rho_D\propto T^2$ Fermi-liquid-like temperature
dependence at low temperatures
\cite{zheng1993,jauho1993,rojo1999}.

One approach in describing Coulomb drag is in terms of a function
$\Gamma_D \left( v_t-v_b \right)$ that gives the rate of change of
momentum per volume of the electron gases due to Coulomb
scattering \cite{jauho1993,rojo1999}, and that to a good
approximation depends on the difference in the drift velocities
$v_t$ and $v_b$ in the top and bottom layers. The equations of
motion for these drift velocities are then given by
\begin{eqnarray}
\label{eq:bilayereomsdriftvs}
  n_t m_e \frac{d v_t}{dt} &=& - \Gamma_D (v_t-v_b) - n_t e E_t - \frac{n_t m_e v_t}{\tau_t}~; \nonumber \\
  n_b m_e \frac{d v_b}{dt} &=& + \Gamma_D (v_t-v_b) - n_b e E_b - \frac{n_b m_e
  v_b}{\tau_b}~.
\end{eqnarray}
Here, $-e$ is the charge of the electron and $m_e$ its mass.
Furthermore, the electronic density and the electric field in the
top (bottom) layer are denoted by $n_t (n_b)$ and $E_t (E_b)$,
respectively. We have also added a scattering rate $1/\tau_t$ and
$1/\tau_b$ for the top and bottom layer, that effectively takes
into account intralayer Coulomb scattering, electron-phonon
interactions, and disorder. In applying the above result to the
situation in Fig.~\ref{fig:dragexpt} we take $v_t=0$ since there
is no current in the top layer. Solving for $v_b$ in the steady
state, using that $\Gamma_D \ll n_b m_e v_b/\tau_b$ because the
interlayer scattering is much weaker than the intralayer
scattering, we find that $v_b = -e E_b \tau_b/m_e$, as usual. In
the linear response regime the drift velocities are small, and we
can use that $\Gamma_D (v) \simeq \Gamma_D ' (0) v$ incorporating
the fact that there is no net momentum transfer if the drift
velocities are equal. We then find that the electric field in the
top layer is $E_t = -\Gamma_D'(0) v_b/en_t$. Using that the
current density in the bottom layer $j_b = -n_b e v_b$, we have
for the drag resistivity
\begin{equation}
\label{eq:dragres}
  \rho_D = \frac{E_t}{j_b} = \frac{\Gamma_D' (0)}{e^2 n_t n_b}~.
\end{equation}
This result shows that the drag resistivity is determined by the
slope of the function $\Gamma_D (v)$ at $v=0$.

In an analogy to Coulomb drag, D'Amico and Vignale proposed spin
Coulomb drag \cite{damico2000}, which was observed by Weber {\it
et al.} \cite{weber2005}. Spin drag, in which the layer degree of
freedom from Coulomb drag is played by the spin of the electrons,
is very similar to Coulomb drag. In this Letter we study spin drag
due to the short-range interatomic interactions in a spin-one Bose
gas in the normal state, and propose an experiment to observe it
making use of so-called spin motive forces. For this system, we
derive equations of motion similar to
Eqs.~(\ref{eq:bilayereomsdriftvs}). The absence of disorder and an
underlying lattice that supports phonons implies that the
analogues of the scattering times $\tau_b,\tau_t$ are infinite.
Nonetheless, we recover a great deal of the phenomenology of
conventional electronic transport. In particular, we define a
transport coefficient analogous to $\rho_D$ which for bosons
becomes large at small temperatures due to Bose enhancement, i.e.,
the enhanced scattering of bosons to states that are already
occupied. In addition, we investigate heating effects and find
that they are completely analogous to the usual Joule heating in
electronic systems.

{\it Ultracold atomic gases and spin motive forces} --- We
consider ultracold atoms with hyperfine spin $F$ in a time and
position dependent magnetic field with a direction given by the
unit vector $\bm{\Omega} (\bx,t)$, such that the Zeeman
interaction reads $H_Z = - \Delta \bm{\Omega} (\bx,t) \cdot {\bf
F}/\hbar$, where ${\bf F}$ are the spin operators and $\Delta$ is
an effective Zeeman splitting energy. If the magnetic-field
direction is varying slowly in space and time, it is convenient to
choose $\bm{\Omega} (\bx,t)$ as the local spin quantization axis.
In this frame of reference spatial and temporal variation of the
magnetic-field direction manifests itself as fictitious, or
fixed-frame, electric and magnetic fields ${\bf E}$ and ${\bf B}$
\cite{stern1992} that are ultimately due to the spin Berry phase
\cite{berry1984}. For atoms with spin projection $m_F$ these are
given by
\begin{eqnarray}
\label{eq:eleandmagnfields}
 E_{m_F,\alpha} &=& m_F \hbar \bm{\Omega} \cdot \left(  \frac{\partial {\bm{\Omega}}}{\partial t}  \times \nabla_\alpha \bm{\Omega}\right)~;
 \nonumber \\
 B_{m_F,\alpha} &=& m_F \hbar \epsilon_{\alpha\beta\gamma} \bm{\Omega} \cdot \left(
 {\nabla_\beta}\bm{\Omega} \times {\nabla_\gamma}\bm{\Omega}
 \right)~,
\end{eqnarray}
where $\hbar$ is Planck's constant, $\epsilon_{\alpha\beta\gamma}$
is the three-dimensional fully antisymmetric Levi-Civita tensor
and a sum over repeated Cartesian indices $\alpha,\beta,\gamma \in
\{x,y,z\}$ is implied. Note that, because the atoms are neutral,
there are no real electromagnetic fields that couple to the atomic
motion. In the context of ferromagnetic metals these fictitious
electric and magnetic fields respectively underlie the phenomena
of spin motive forces induced by moving domain walls and the
topological Hall effect, both of which have been observed very
recently \cite{yang2009,neubauer2009}. In the context of cold
atoms, the Aharonov-Bohm phase due to the fictitious magnetic
field, in combination with phase coherence, has been used to
imprint coreless vortices on $F=1$ spinor Bose-Einstein
condensates \cite{isoshima2000,leanhardt2003}. For the existence
of the fictitious electric and magnetic fields phase coherence is,
however, not required \cite{stern1992} and we can focus instead on
the semi-classical regime using the equation of motion
\begin{eqnarray}
\label{eq:semicleqofm}
  m \frac{ d {\bf v}_{m_F}}{dt} =  {\bf E}_{m_F} + {\bf v}_{m_F} \times
  {\bf B}_{m_F}~,
\end{eqnarray}
for an atom with velocity ${\bf v}_{m_F}$ and spin projection
$m_F$.

\begin{figure}
\vspace{-0.5cm} \centerline{\epsfig{figure=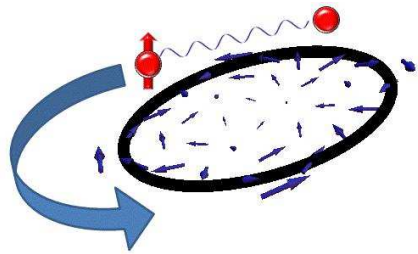,width=6cm}}
 \caption{(Color online) Illustration of spin drag in a toroidal trap. The atoms with $m_F=1$ (red circle with arrow) are accelerated by
 the motive force due to the time-dependent quadrupole field (blue arrows) of the Ioffe-Pritchard trap. Due to interactions the atoms with $m_F=0$
 are dragged along.}
 \label{fig:torus}
\end{figure}

The specific geometry we consider is illustrated in
Fig.~\ref{fig:torus} and consists of a toroidal trap with radius
$R$ and effective cross section area $A$ in the transverse
direction, created by a rapidly-moving laser beam
\cite{henderson2009}. To implement the fictitious electric field
we superpose a Ioffe-Pritchard magnetic trap. Fictitious electric
fields along the torus are achieved by varying the axial bias
field of the Ioffe-Pritchard trap, so that \cite{leanhardt2003}
\begin{equation}
  \bm{\Omega} (\phi) = \Omega_z (t) \hat z + \sqrt{1-\Omega_z^2 (t)}
  \left[ \hat r \cos 2 \phi   - \hat \phi \sin 2 \phi  \right]~,
\end{equation}
in cylindrical coordinates $(r,\phi,z)$. Using
Eq.~(\ref{eq:eleandmagnfields}) we then find that
\begin{equation}
\label{eq:electricfield}
  {\bf E}_{m_F} =  2 \hbar m_F \frac{1}{R} \frac{d \Omega_z (t)}{dt}  \hat \phi~,
\end{equation}
and ${\bf B}_{m_F}=0$. The adiabatic approximation that leads to
the above holds when the timescale $T_0$ on which the direction of
the external magnetic field is changed is much larger than the
spin precession time $\hbar/\Delta$. Furthermore, this spin
precession time should be smaller than the time it takes the atoms
to encircle the torus. Since $\Delta$ is a large energy scale,
these conditions are easily satisfied.

{\it Spin drag} --- We now specifically consider noncondensed
bosonic atoms with $F=1$, e.g., sodium atoms. Furthermore, to
study spin drag we consider the case that the trap is loaded with
equal densities of atoms in spin state $|1\rangle$ with $m_F=+1$
and $|0\rangle$ with $m_F=0$. According to
Eq.~(\ref{eq:electricfield}) the atoms in spin state $|1\rangle$
then feel a fictitious electric field $E_1 \equiv E$ along the
torus which accelerates them.  The atoms in the $|0\rangle$ state
feel no fictitious electric field but may accelerate due to spin
drag, i.e., due to collisions with the other atoms.

To investigate the spin drag quantitatively, we use an effective
one-dimensional Boltzmann equation for the distribution function
$f_1 (k,t)$ and $f_0 (k,t)$ of the $m_F=1$ and $m_F=0$ atoms, with
$\hbar k$ the momentum along the torus, given by
\begin{widetext}
\begin{eqnarray}
\label{eq:qbe}
  && \frac{\partial f_1 (k,t)}{\partial t} + \frac{E}{\hbar}
  \frac{\partial f_1}{\partial k} = \Gamma_{\rm coll}(k) \equiv \frac{2\pi}{\hbar}
  \left(T^{2B}_{01}\right)^2
  \int \frac{dk_2}{2\pi} \int \frac{dk_3}{2\pi} \int \frac{dk_4}{2\pi}
  (2\pi) \delta (k+k_2-k_3-k_4) \delta
  \left(\epsilon_{k}+\epsilon_{k_2}-\epsilon_{k_3}-\epsilon_{k_4}\right)\nonumber
  \\
  && \times \left\{
    \left[ 1+f_1 (k,t) \right] \left[ 1+f_0 (k_2,t) \right]f_1
    (k_3,t) f_0 (k_4,t)
    -f_1 (k,t) f_0 (k_2,t) \left[1+f_1 (k_3,t)  \right] \left[ 1+f_0 (k_4,t)  \right]
  \right\}~,
\end{eqnarray}
\end{widetext}
and where the equation of motion for $f_0 (k,t)$ follows by taking
$E=0$ and interchanging $f_0 (k,t)$ and $f_1 (k,t)$. The
interspecies collisions are determined by the two-body T-matrix
$T^{2B}_{01} = 4 \pi a \hbar^2/m A$, with $a$ the scattering
length for collisions of atoms between atoms in states $|1\rangle$
and $|0\rangle$, and $m$ the atomic mass. The single-particle
dispersion is $\epsilon_k=\hbar^2 k^2/2m$. On the right-hand side
we have ignored intra-spin-species collisions which tend to
restore local equilibrium and are zero in the approximations
outlined below. Also note that, contrary to electronic transport
in solid-state physics, there are no terms corresponding to
elastic or electron-phonon collisions because in cold-atom systems
there is no extrinsic disorder or an underlying ionic lattice.

\begin{figure}
\vspace{-0.5cm} \centerline{\epsfig{figure=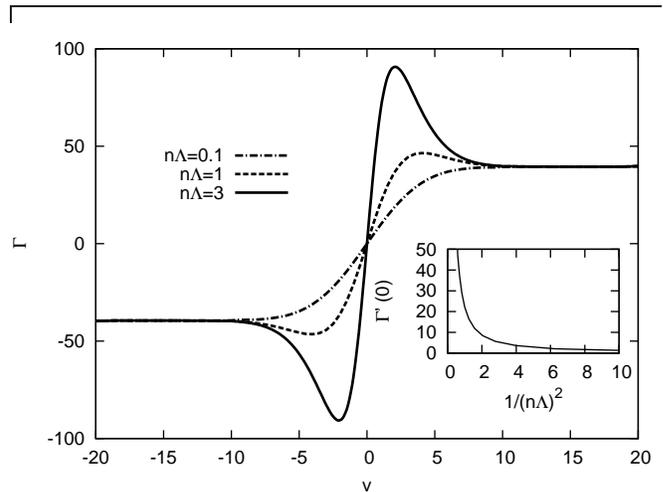}}
 \caption{Rate of momentum transfer $\Gamma$ in units of $A^2 m/(2 a \hbar n)^2$. The difference of drift velocities is in units of $\hbar/m\Lambda$.
 The inset shows the derivative of $\Gamma$ at $v=0$.}
 \label{fig:gamma_v}
\end{figure}

Since the intraspecies collision enforce local equilibrium for
each spin species we use a Bose-Einstein distribution function
with nonzero drift velocity as an {\it ansatz} to solve the above
equation. Specifically, we take $f_1 (k,t) = N_B (\epsilon_{k-mv_1
(t)/\hbar})$ and $f_0 (k,t) = N_B (\epsilon_{k-mv_0 (t)/\hbar})$,
with $N_B (\epsilon)=[ e^{\beta_T (\epsilon -\mu)}-1]^{-1}$ the
Bose-Einstein distribution function at chemical potential $\mu$
and inverse thermal energy $\beta_T=1/k_B T$. In first instance we
take the temperature constant in time. The time dependence of the
chemical potential is determined by the conservation of the number
of atoms in each spin state and is left implicit. From the
Boltzmann equation we then find that, cf.
Eqs.~(\ref{eq:bilayereomsdriftvs}),
\begin{eqnarray}
\label{eq:eomsforatoms}
  n m \frac{d v_1}{dt} &=& n E + \Gamma (v_0-v_1)~;\nonumber \\
  n m \frac{d v_0}{dt} &=& -\Gamma (v_0-v_1)~,
\end{eqnarray}
where $n$ is the one-dimensional density, and the function that
determines the rate of momentum transfer from species $|1\rangle$
to $|0\rangle$ is found from Eq.~(\ref{eq:qbe}) as
\begin{equation}
\label{eq:momtransrate} \Gamma (v_0-v_1) = -  \int \frac{dk}{2\pi}
\Gamma_{\rm coll} (k) \hbar k~,
\end{equation}
with the right-hand side evaluated using the shifted Bose-Einstein
distribution functions. In Fig.~\ref{fig:gamma_v} we plot this
function for various values of the degeneracy parameter
$n\Lambda$, with $\Lambda= \sqrt{2\pi\hbar^2/mk_BT}$ the deBroglie
wave length. We find that in the classical limit $n\Lambda \to 0$
it is given by $\Gamma (v) = (4 \pi a \hbar n)^2 {\rm Erf }
(m\Lambda v/\hbar) /A^2m$. For increasing degeneracy $\Gamma (v)$
develops local maxima and minima at small $|v_{\rm max}|$ which
are due to Bose enhancement of interspecies scattering.

From the equations of motion in Eq.~(\ref{eq:eomsforatoms}) we see
that the sum of drift velocities increases indefinitely. The
relative drift velocity $v=v_1-v_0$ can approach a steady state,
provided the motive force $E$ is not too large. That is, from
Eq.~(\ref{eq:eomsforatoms}) we find that if $n E \leq 2\Gamma
(v_{\rm max})$ the system approaches a steady state where
$dv/dt=0$. In the linear-response regime $E$ and $v$ are small and
we have that $\Gamma (v_0-v_1) \simeq \Gamma' (0) (v_0-v_1)$.
Introducing the relative-momentum particle current $j = n
(v_1-v_0)$, we have in linear response that $v_1-v_0 =
nE/2\Gamma'(0)$. From this we define in the linear-response regime
a resistivity $\rho \equiv E/j=2\Gamma'(0)/n^2$, that is analogous
to the drag resistivity in Eq.~(\ref{eq:dragres}). For fermionic
atoms this resistivity would vanish at small temperatures. For
bosons it becomes larger due to Bose enhancement. This is further
illustrated in the inset of Fig.~\ref{fig:gamma_v} which shows
$\Gamma' (0)$ as a function of $1/(n\Lambda)^2 \propto T$.

The fact that the total kinetic energy of the system is increasing
suggests that beyond-linear-response effects, such as heating, may
be important. To investigate these, we have to solve
Eqs.~(\ref{eq:eomsforatoms}) coupled to an equation for the
temperature. This equation is most easily derived by considering
the total energy $U = \int (f_1+f_0) \epsilon_k dk/2\pi$. We
evaluate this energy within our {\it ansatz} of Bose-Einstein
distribution functions with nonzero drift velocities and time-dependent temperature $T(t)$ in this case. Using
the Boltzmann equation in Eq.~(\ref{eq:qbe}) and
Eqs.~(\ref{eq:eomsforatoms}), we find that $
  dQ/dt = n (v_1-v_0) \Gamma (v_1-v_0)$,
where $Q \equiv U-mn(v_1^2+v_0^2)/2$. This energy is determined by
the spread in velocities in the gas of atoms and is therefore a
measure for its temperature. Evaluating the above using the
linear-response expression for the difference in drift velocities,
we find that $dQ/dt = n \rho j^2/2$, which is analogous to Ohmic
heating in electronic systems.

\begin{figure}
\vspace{-0.5cm} \centerline{\epsfig{figure=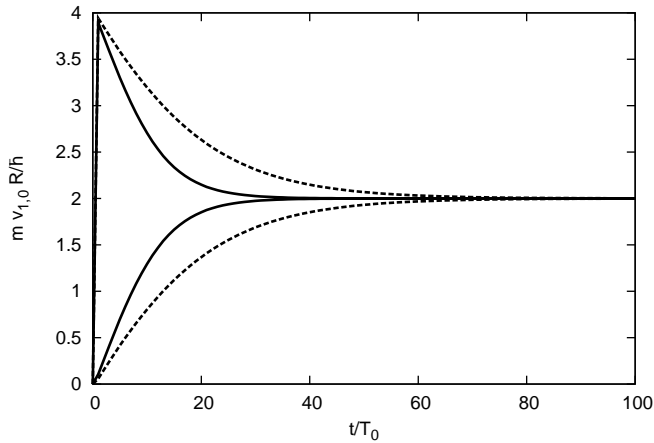}}
 \caption{Drift velocities of as a function of time, for $n\Lambda=45$ (solid lines) and $n\Lambda=30$ (dashed lines). The upper and lower curve
 correspond to $v_1$ and $v_0$, respectively.}
 \label{fig:velos}
\end{figure}

We go beyond linear response by solving the equation for $dQ/dt$
coupled to Eqs.~(\ref{eq:eomsforatoms}).  We consider the case
that the axial magnetic field of the Ioffe-Pritchard trap is
inverted in a time $T_0$, so that $\Omega_z (t) =(2t-1)/T_0$ for
$0<t<T_0$, and constant for $t>T_0$.  This implies via
Eq.~(\ref{eq:electricfield}) that ${\rm E}_{m_F} =
4\hbar\hat\phi/RT_0$ for $0<t<T_0$ and zero for $t>T_0$. We
consider specifically $^{23}$Na atoms. As parameters we take
$T_0=10$ ms \cite{leanhardt2003}, $R=5$ $\mu$m, $T=400$ nK. For
the one-dimensional density we take $n=10^{12}$ cm$^{-3}$ $\times
A$, with $A=\pi$ ($5 \mu$m)$^2$ \cite{ryu2007}. For these
parameters $n\Lambda=45$. The result is shown in
Fig.~\ref{fig:velos}, together with the result for $n\Lambda=30$.
We find that heating effects are negligible on the time scale that
is shown. For each pair of curves the upper one corresponds to
$v_1 (t)$, which, due to acceleration by the motive force,
acquires the value $v_1 \simeq 4 \hbar/mR$ in the time $T_0$. The
lower curve corresponds to $v_0$ which starts at $v_0 (0)=0$. Due
to the spin drag, the latter velocity also becomes nonzero, which
can be experimentally measured by studying the momentum
distribution after expansion. Note that the drag effect is larger
for larger $n\Lambda$ due to the Bose enhancement.

{\it Discussion and conclusions} --- There are other experimental
setups possible to observe spin drag effects. For example, a
cigar-shaped optical trap together with a magnetic field gradient
in the axial direction also leads to relative motion of the two
spin species. We note that the hydrodynamic regime, where
spin-drag effects should be large, has been realized recently in
such cigar-shaped systems \cite{vanderstam2007}. Another
possibility is using a Raman transition to convert a fraction of
the atoms of one spin species to another, and to set them into
motion with a velocity determined by the recoil energy of the
two-photon process. Such an experiment has already been performed
with Bose-Einstein condensates \cite{chikkatur2000}. However, to
study spin drag, and other analogues of electronic transport, the
noncondensed case is more suitable because the incoherent
collisions of the thermal atoms, rather than the coherent
interactions in a Bose-Einstein condensate, are analogous to the
collisions of the electrons.

Yet another experimental possibility is to use a
sinusoidally-varying axial bias field of the Ioffe-Pritchard trap.
This would lead to an AC electric field, and the possibility to
measure the frequency dependence of the transport coefficient
$\rho$. Other interesting generalizations of the present work are
including mesoscopic phase-coherence effects, and effects of
critical fluctuations. Drag effects can also be measured in Fermi
gases, leading to another way to probe the many-body physics of
these systems.

We thank Peter van der Straten, Robert Meppelink, and Johnny
Vogels for pointing out to us various experimental
possibilities to measure the spin drag, and David Ritchie for allowing us to use Fig.~\ref{fig:dragexpt}. This work was supported by
the Stichting voor
Fundamenteel Onderzoek der Materie (FOM), the Netherlands Organization for Scientific Research (NWO), and by
the European Research Council (ERC) under the Seventh Framework
Program (FP7).


\vspace{-0.5cm}
\begin{thebibliography}{17}
\expandafter\ifx\csname
natexlab\endcsname\relax\def\natexlab#1{#1}\fi
\expandafter\ifx\csname bibnamefont\endcsname\relax
  \def\bibnamefont#1{#1}\fi
\expandafter\ifx\csname bibfnamefont\endcsname\relax
  \def\bibfnamefont#1{#1}\fi
\expandafter\ifx\csname citenamefont\endcsname\relax
  \def\citenamefont#1{#1}\fi
\expandafter\ifx\csname url\endcsname\relax
  \def\url#1{\texttt{#1}}\fi
\expandafter\ifx\csname
urlprefix\endcsname\relax\def\urlprefix{URL }\fi
\providecommand{\bibinfo}[2]{#2}
\providecommand{\eprint}[2][]{\url{#2}}

\bibitem[{\citenamefont{Rammer}(1998)}]{rammerbook}
\bibinfo{author}{\bibfnamefont{J.}~\bibnamefont{Rammer}},
  \emph{\bibinfo{title}{Quantum Transport Theory}}
  (\bibinfo{publisher}{Westview Press}, \bibinfo{address}{Boulder, CO},
  \bibinfo{year}{1998}), \bibinfo{edition}{1st} ed.

\bibitem[{\citenamefont{Gramila et~al.}(1991)\citenamefont{Gramila, Eisenstein,
  MacDonald, Pfeiffer, and West}}]{gramila1991}
\bibinfo{author}{\bibfnamefont{T.~J.} \bibnamefont{Gramila}},
  \bibinfo{author}{\bibfnamefont{J.~P.} \bibnamefont{Eisenstein}},
  \bibinfo{author}{\bibfnamefont{A.~H.} \bibnamefont{MacDonald}},
  \bibinfo{author}{\bibfnamefont{L.~N.} \bibnamefont{Pfeiffer}},
  \bibnamefont{and} \bibinfo{author}{\bibfnamefont{K.~W.} \bibnamefont{West}},
  \bibinfo{journal}{Phys. Rev. Lett.} \textbf{\bibinfo{volume}{66}},
  \bibinfo{pages}{1216} (\bibinfo{year}{1991}).

\bibitem[{\citenamefont{Zheng and MacDonald}(1993)}]{zheng1993}
\bibinfo{author}{\bibfnamefont{L.}~\bibnamefont{Zheng}} \bibnamefont{and}
  \bibinfo{author}{\bibfnamefont{A.~H.} \bibnamefont{MacDonald}},
  \bibinfo{journal}{Phys. Rev. B} \textbf{\bibinfo{volume}{48}},
  \bibinfo{pages}{8203} (\bibinfo{year}{1993}).

\bibitem[{\citenamefont{Jauho and Smith}(1993)}]{jauho1993}
\bibinfo{author}{\bibfnamefont{A.-P.} \bibnamefont{Jauho}} \bibnamefont{and}
  \bibinfo{author}{\bibfnamefont{H.}~\bibnamefont{Smith}},
  \bibinfo{journal}{Phys. Rev. B} \textbf{\bibinfo{volume}{47}},
  \bibinfo{pages}{4420} (\bibinfo{year}{1993}).

\bibitem[{\citenamefont{Rojo}(1999)}]{rojo1999}
\bibinfo{author}{\bibfnamefont{A.~G.} \bibnamefont{Rojo}}, \bibinfo{journal}{J.
  Phys: Cond. Matt.} \textbf{\bibinfo{volume}{11}}, \bibinfo{pages}{R31}
  (\bibinfo{year}{1999}).

\bibitem[{\citenamefont{D\char39{}Amico and Vignale}(2000)}]{damico2000}
\bibinfo{author}{\bibfnamefont{I.}~\bibnamefont{D\char39{}Amico}}
  \bibnamefont{and} \bibinfo{author}{\bibfnamefont{G.}~\bibnamefont{Vignale}},
  \bibinfo{journal}{Phys. Rev. B} \textbf{\bibinfo{volume}{62}},
  \bibinfo{pages}{4853} (\bibinfo{year}{2000}).

\bibitem[{\citenamefont{Weber et~al.}(2005)\citenamefont{Weber, Gedik, Moore,
  Orenstein, Stephens, and Awschalom}}]{weber2005}
\bibinfo{author}{\bibfnamefont{C.}~\bibnamefont{Weber}},
  \bibinfo{author}{\bibfnamefont{N.}~\bibnamefont{Gedik}},
  \bibinfo{author}{\bibfnamefont{J.}~\bibnamefont{Moore}},
  \bibinfo{author}{\bibfnamefont{J.}~\bibnamefont{Orenstein}},
  \bibinfo{author}{\bibfnamefont{J.}~\bibnamefont{Stephens}}, \bibnamefont{and}
  \bibinfo{author}{\bibfnamefont{D.}~\bibnamefont{Awschalom}},
  \bibinfo{journal}{Nature} \textbf{\bibinfo{volume}{437}},
  \bibinfo{pages}{1330} (\bibinfo{year}{2005}).

\bibitem[{\citenamefont{Stern}(1992)}]{stern1992}
\bibinfo{author}{\bibfnamefont{A.}~\bibnamefont{Stern}},
  \bibinfo{journal}{Phys. Rev. Lett.} \textbf{\bibinfo{volume}{68}},
  \bibinfo{pages}{1022} (\bibinfo{year}{1992}).

\bibitem[{\citenamefont{Berry}(1984)}]{berry1984}
\bibinfo{author}{\bibfnamefont{M.~V.} \bibnamefont{Berry}},
  \bibinfo{journal}{Proc. R. Soc. London A} \textbf{\bibinfo{volume}{392}},
  \bibinfo{pages}{45} (\bibinfo{year}{1984}).

\bibitem[{\citenamefont{Yang et~al.}(2009)\citenamefont{Yang, Beach, Knutson,
  Xiao, Niu, Tsoi, and Erskine}}]{yang2009}
\bibinfo{author}{\bibfnamefont{S.~A.} \bibnamefont{Yang}},
  \bibinfo{author}{\bibfnamefont{G.~S.~D.} \bibnamefont{Beach}},
  \bibinfo{author}{\bibfnamefont{C.}~\bibnamefont{Knutson}},
  \bibinfo{author}{\bibfnamefont{D.}~\bibnamefont{Xiao}},
  \bibinfo{author}{\bibfnamefont{Q.}~\bibnamefont{Niu}},
  \bibinfo{author}{\bibfnamefont{M.}~\bibnamefont{Tsoi}}, \bibnamefont{and}
  \bibinfo{author}{\bibfnamefont{J.~L.} \bibnamefont{Erskine}},
  \bibinfo{journal}{Phys. Rev. Lett.} \textbf{\bibinfo{volume}{102}},
  \bibinfo{eid}{067201} (\bibinfo{year}{2009}).

\bibitem[{\citenamefont{Neubauer et~al.}(2009)\citenamefont{Neubauer,
  Pfleiderer, Binz, Rosch, Ritz, Niklowitz, and B\"{o}ni}}]{neubauer2009}
\bibinfo{author}{\bibfnamefont{A.}~\bibnamefont{Neubauer}},
  \bibinfo{author}{\bibfnamefont{C.}~\bibnamefont{Pfleiderer}},
  \bibinfo{author}{\bibfnamefont{B.}~\bibnamefont{Binz}},
  \bibinfo{author}{\bibfnamefont{A.}~\bibnamefont{Rosch}},
  \bibinfo{author}{\bibfnamefont{R.}~\bibnamefont{Ritz}},
  \bibinfo{author}{\bibfnamefont{P.~G.} \bibnamefont{Niklowitz}},
  \bibnamefont{and} \bibinfo{author}{\bibfnamefont{P.}~\bibnamefont{B\"{o}ni}},
  \bibinfo{journal}{Phys. Rev. Lett.} \textbf{\bibinfo{volume}{102}},
  \bibinfo{eid}{186602} (\bibinfo{year}{2009}).

\bibitem[{\citenamefont{Isoshima et~al.}(2000)\citenamefont{Isoshima, Nakahara,
  Ohmi, and Machida}}]{isoshima2000}
\bibinfo{author}{\bibfnamefont{T.}~\bibnamefont{Isoshima}},
  \bibinfo{author}{\bibfnamefont{M.}~\bibnamefont{Nakahara}},
  \bibinfo{author}{\bibfnamefont{T.}~\bibnamefont{Ohmi}}, \bibnamefont{and}
  \bibinfo{author}{\bibfnamefont{K.}~\bibnamefont{Machida}},
  \bibinfo{journal}{Phys. Rev. A} \textbf{\bibinfo{volume}{61}},
  \bibinfo{pages}{063610} (\bibinfo{year}{2000}).

\bibitem[{\citenamefont{Leanhardt et~al.}(2003)\citenamefont{Leanhardt, Shin,
  Kielpinski, Pritchard, and Ketterle}}]{leanhardt2003}
\bibinfo{author}{\bibfnamefont{A.~E.} \bibnamefont{Leanhardt}},
  \bibinfo{author}{\bibfnamefont{Y.}~\bibnamefont{Shin}},
  \bibinfo{author}{\bibfnamefont{D.}~\bibnamefont{Kielpinski}},
  \bibinfo{author}{\bibfnamefont{D.~E.} \bibnamefont{Pritchard}},
  \bibnamefont{and} \bibinfo{author}{\bibfnamefont{W.}~\bibnamefont{Ketterle}},
  \bibinfo{journal}{Phys. Rev. Lett.} \textbf{\bibinfo{volume}{90}},
  \bibinfo{pages}{140403} (\bibinfo{year}{2003}).

\bibitem[{\citenamefont{Henderson et~al.}(2009)\citenamefont{Henderson, Ryu,
  MacCormick, and Boshier}}]{henderson2009}
\bibinfo{author}{\bibfnamefont{K.}~\bibnamefont{Henderson}},
  \bibinfo{author}{\bibfnamefont{C.}~\bibnamefont{Ryu}},
  \bibinfo{author}{\bibfnamefont{C.}~\bibnamefont{MacCormick}},
  \bibnamefont{and} \bibinfo{author}{\bibfnamefont{M.~G.}
  \bibnamefont{Boshier}}, \bibinfo{journal}{New J. Phys.}
  \textbf{\bibinfo{volume}{11}}, \bibinfo{pages}{043030}
  (\bibinfo{year}{2009}).

\bibitem[{\citenamefont{Ryu et~al.}(2007)\citenamefont{Ryu, Andersen,
  Clad\'{e}, Natarajan, Helmerson, and Phillips}}]{ryu2007}
\bibinfo{author}{\bibfnamefont{C.}~\bibnamefont{Ryu}},
  \bibinfo{author}{\bibfnamefont{M.~F.} \bibnamefont{Andersen}},
  \bibinfo{author}{\bibfnamefont{P.}~\bibnamefont{Clad\'{e}}},
  \bibinfo{author}{\bibfnamefont{V.}~\bibnamefont{Natarajan}},
  \bibinfo{author}{\bibfnamefont{K.}~\bibnamefont{Helmerson}},
  \bibnamefont{and} \bibinfo{author}{\bibfnamefont{W.~D.}
  \bibnamefont{Phillips}}, \bibinfo{journal}{Phys. Rev. Lett.}
  \textbf{\bibinfo{volume}{99}}, \bibinfo{eid}{260401} (\bibinfo{year}{2007}).

\bibitem[{\citenamefont{van~der Stam et~al.}(2007)\citenamefont{van~der Stam,
  Meppelink, Vogels, and van~der Straten}}]{vanderstam2007}
\bibinfo{author}{\bibfnamefont{K.~M.~R.} \bibnamefont{van~der Stam}},
  \bibinfo{author}{\bibfnamefont{R.}~\bibnamefont{Meppelink}},
  \bibinfo{author}{\bibfnamefont{J.~M.} \bibnamefont{Vogels}},
  \bibnamefont{and} \bibinfo{author}{\bibfnamefont{P.}~\bibnamefont{van~der
  Straten}}, \bibinfo{journal}{Phys. Rev. A} \textbf{\bibinfo{volume}{75}},
  \bibinfo{eid}{031602(R)} (\bibinfo{year}{2007}).

\bibitem[{\citenamefont{Chikkatur et~al.}(2000)\citenamefont{Chikkatur,
  G\"orlitz, Stamper-Kurn, Inouye, Gupta, and Ketterle}}]{chikkatur2000}
\bibinfo{author}{\bibfnamefont{A.~P.} \bibnamefont{Chikkatur}},
  \bibinfo{author}{\bibfnamefont{A.}~\bibnamefont{G\"orlitz}},
  \bibinfo{author}{\bibfnamefont{D.~M.} \bibnamefont{Stamper-Kurn}},
  \bibinfo{author}{\bibfnamefont{S.}~\bibnamefont{Inouye}},
  \bibinfo{author}{\bibfnamefont{S.}~\bibnamefont{Gupta}}, \bibnamefont{and}
  \bibinfo{author}{\bibfnamefont{W.}~\bibnamefont{Ketterle}},
  \bibinfo{journal}{Phys. Rev. Lett.} \textbf{\bibinfo{volume}{85}},
  \bibinfo{pages}{483} (\bibinfo{year}{2000}).

\end{thebibliography}
\end{document}